\documentclass[]{interact}

\usepackage{epstopdf}
\usepackage[caption=false]{subfig}
\usepackage[doublespacing]{setspace}
\usepackage[colorlinks=true,citecolor=blue]{hyperref}%
\usepackage{verbatim}
\usepackage{soul}
\usepackage[normalem]{ulem}
\usepackage{color}
\usepackage[numbers,sort&compress]{natbib}
\bibpunct[, ]{[}{]}{,}{n}{,}{,}
\makeatletter
\def\NAT@def@citea{\def\@citea{\NAT@separator}}
\makeatother

\theoremstyle{plain}

\theoremstyle{definition}

\theoremstyle{remark}

\begin{document}


\title{Precursors to plastic failure in a numerical simulation of CuZr metallic glass}
\author{
\name{Matias Sepulveda--Macias\textsuperscript{a}$^{,*}$\thanks{$^*$~Corresponding author. Email: msepulvedamacias@ing.uchile.cl}, Gonzalo Gutierrez\textsuperscript{b}\thanks{G. Gutierrez. Email: gonzalogutierrez@uchile.cl} and Fernando Lund\textsuperscript{a}\thanks{F. Lund. Email: flund@dfi.uchile.cl}}
\affil{\textsuperscript{a}Departamento de F\'isica, Facultad de Ciencias F\'isicas y Matem\'aticas, Universidad de Chile, Santiago, Chile; \textsuperscript{b}Grupo de NanoMateriales, Departamento de F\'isica, Facultad de Ciencias, Universidad de Chile, Casilla 653, Santiago, Chile.}
}

\maketitle

\begin{abstract}
We deform, in pure shear, a thin sample of Cu$_{50}$Zr$_{50}$ metallic glass using a molecular dynamics simulation up to, and including, failure. The experiment is repeated ten times in order to have average values and standard deviations. Although failure occurs at the same value of the externally imposed strain for the ten samples, there is significant sample-to-sample variation in the specific microscopic material behavior.  Failure can occur along one, two, or three planes, located at the boundaries of previously formed shear bands. These shear bands form shortly before failure. However, well before their formation and at external strains where plastic deformation just begins to be significant, non--affine displacement organizes itself along localized bands. The shear bands subsequently form at the edges of these non--affine--displacement--bands, and present an alternating rotation-quadrupole structure, as found previously by \c{S}opu~\emph{et al.}~\cite{Sopu2017}  in the case of a notched sample loaded in tension. The thickness of shear bands is roughly determined by the available plastic energy. The onset of shear banding is accompanied by a sharp increase in the rate of change of the rotation angle localization, the strain localization, and the non--affine square displacement.

\end{abstract}


\section{Introduction}
Metallic glasses (MGs) and, particularly, bulk metallic glasses (BMGs) have been in existence since 
1960~\cite{Klement1960} and since 1993~\cite{Peker1993,Inoue1993} respectively. They share a number 
of properties with oxide, and other, glasses, most notably the lack of long--range translational order. 
They differ, in that they are kept together not by covalent or ionic bonding but by metallic bonding, 
which does not have a preferred direction. This is an important ingredient leading to the fact that, 
in order to achieve a glassy state, melts have to be cooled quite rapidly in order to avoid 
crystallization~\cite{suryanarayana2011}. The fact that BMGs can be mold--cast has enabled a number 
of applications for the fabrication of parts that need to have a prescribed shape and where a 
combination of high strength and low stiffness is useful. The fast cooling rate that is needed, 
however, severely limits their size.

The mechanical behavior of BMGs is a noteworthy characteristic  that renders them unique~\cite{Hufnagel2016}. 
At very low strains, their mechanical response is linear, similar to their crystalline counterparts. As the 
loading is increased however, crystalline metals develop dislocations and undergo plastic deformation. MGs, 
not having a crystal lattice, do not have this property and are, to a first approximation, brittle. This fact 
points to a fundamental question: how is one to link the microstructure of a MG with its macroscopic mechanical 
response? In crystals, this link is provided by the dislocations. Plastic flow does occur, though, in MGs, 
although to a much more limited extent than in crystals, and it is localized within thin shear bands (SBs)~\cite{Greer2013}. 
The appearance of SBs is often a signal of imminent failure, so it is critical to achieve as complete an 
understanding as possible of their behavior. Lacking the analytical tools that would be enabled by the existence 
of a crystalline structure, numerical simulation is the tool of choice to bridge the gap between the vastly 
different length scales that span the atomic to macroscopic domains.

Among the many existing MGs, one that has been the object of many studies, both experimental and numerical, 
is CuZr, with a variety of relative Cu contents, and with the possible addition of other elements such as 
Al~\cite{MeiBo2004,Tercini2018}. This interest stems from their superior glass forming ability~\cite{Xu2004}, 
sensitivity of mechanical behavior to composition~\cite{Kumar2007}, and the existence of reliable interatomic potentials 
for numerical simulations~\cite{Mendelev2007,Mendelev2009,Borovikov2016}. In this sense, it has been possible 
to establish, for example, that the birth of multiple shear bands, induced by the presence of a crystalline 
phase in the glass matrix, avoids the fragile fracture and gives way to the formation of a plastic regime 
in the material~\cite{Sopu2015}. In addition, the study of the nature of the shear transformation zones (STZs) 
and the interaction between them for the formation of SBs has been intensified. 

Recently, \c{S}opu~\emph{et al.}~\cite{Sopu2017} have studied the growth of a SB in a notched specimen 
loaded in tension using molecular dynamics (MD) simulations. They have shown that a shear band is initiated, as expected,
at the stress concentration due to the notch and propagates, also as expected, along a plane of maximum resolved shear stress. 
This propagation proceeds through a successive generation of vortex--like structures (associated with non--affine 
displacement (NAD)), and STZs. A study of the non--affine displacement field inside the SBs revealed the existence 
of these vortex--like structures, which represent the medium that transmits the distortion between STZs and that also 
have a high concentration of stress. They are, therefore, able to store enough energy to activate the next STZ. 
It was concluded that an adequate control over these ``STZ--vortex'' regions could lead to the control of the SB 
dynamics, perhaps avoiding instability.

Following previous work of us~\cite{SepulvedaMacias2016}, in the present work we use MD simulations to study the generation of SBs in a Cu$_{50}$Zr$_{50}$ MG, paying special attention to the evolution of non--affine displacement. Rather than a notched specimen in tension, we consider a volume--conserving simple shear deformation in a sample that does not initially have any externally introduced flaws. Nevertheless, the mechanism of Ref.~\cite{Sopu2017} appears to be, broadly, at work here as well, although in a different geometrical setting. In order to have proper averages and standard deviations of what is an essentially random behavior, we have repeated each numerical experiment ten times. We have studied the behavior of the NAD field, the local von Mises strain, local rotation angle and the local five--fold symmetry.

\section{Simulation details}\label{simulacion}

Molecular dynamics simulations for the Cu$_{50}$Zr$_{50}$
were carried out using LAMMPS software~\cite{Plimpton19951}. An embedded 
atom model (EAM) potential developed by Cheng~\emph{et al.}~\cite{Cheng2011} 
was adopted to describe the interatomic interactions, which has been widely used to 
model Cu-Zr binary systems.

We build ten simulation samples, all of them consisting of 580 800 atoms, with dimensions 
of $80\times40\times2$~nm$^3$. The reason for considering ten samples is discussed at 
the end of the paragraph. The reason for choosing one dimension much smaller than the other 
two is to facilitate the visualization of SBs. Each sample was made in the following way: 
We start with a FCC copper structure of 145 200 atoms and replace 50\% of atoms by zirconium 
atoms at random positions. These systems are then heated from 300 K in the NPT ensemble for 
2 ns up to 2 200 K and keeping a constant 0 GPa pressure. The integration timestep was set at 1 fs. 
Then, the molten metal is cooled down to 10 K following a procedure described by 
Wang~\emph{et al.}~\cite{Wang2012}. The estimated cooling rate is $10^{12}$~K$~\text{s}^{-1}$. 
This 145 200 atoms glass is replicated in $x$ and $y$ direction, followed by an equilibration  
in the NPT ensemble in order to remove eventual problems in the replication process, as 
described in~\cite{Arman2010}. 
Finally we let the system evolve in the NVE ensemble during 100 ps with a final minimization to 
ensure that all atomic forces were under $10^{-4}$~eV~\AA$^{-1}$, thus obtaining well--equilibrated 
metallic glass samples. Since each material sample is a single realization of a random material, 
we have repeated the procedure ten times in order to have conclusions that are reasonably 
representative of average properties and not just a statistical exception.

On these well--equilibrated samples we applied an engineering shear strain of $\dot\gamma=5\times10^8$~s$^{-1}$, up to 
a maximum macroscopic deformation of $\gamma = 0.20$. This is applied to the simulation cell in the $xy$ direction, using the NVE ensemble, 
and periodic boundary conditions (PBC) were set in all directions. When $\gamma = 0.138\pm0.006$, failure occurs. This is ascertained by the observed  macroscopic displacement of one part of the sample relative to the other along a shear band. Although we continue the simulation, something that periodic boundary conditions allow us to do, up to  $\gamma = 0.20$ in order to ensure numerical stability, we are interested in the physics of the atomic processes leading up to failure. 

\subsection{Analytical tools}

Several diagnostic tools were employed to analyze our simulations. The first is the von Mises strain invariant, commonly known 
as local atomic shear strain, $\eta^{\text{Mises}}$, proposed by Shimizu~\emph{et al.}~\cite{Shimizu2007}. This parameter is 
calculated by comparing two atomic configurations: current and reference, and computing the per--particle strain tensor 
$\eta^{i}_{\alpha \beta}$, as defined by the relative translation of its nearest neighbors: 
\begin{equation}\label{shearstrain}
\eta_i^{\text{Mises}} = \sqrt{(\eta_{yz}^{i})^2 + (\eta_{xz}^{i})^2 + (\eta_{xy}^{i})^2 + \frac{(\eta^{i}_{yy}-\eta^{i}_{zz})^2 + (\eta^{i}_{xx}-\eta^{i}_{zz})^2 + (\eta^{i}_{xx}-\eta^{i}_{yy})^2}{6}}.
\end{equation}
A global parameter that is a useful indicator of the overall sample behavior is the standard deviation of $\eta^{\text{Mises}}$ from its mean, the 
``degree of strain localization''~\cite{Cheng2009-vonmises} :
\begin{equation}
\psi = \sqrt{\frac1N \sum_{i=1}^N\left(\eta_i^{\text{Mises}} - \eta^{\text{Mises}}_{\text{ave}}\right)^2}\ ,
\label{eq:psi}
\end{equation}
where $\eta_{\text{ave}}^{\text{Mises}}$ corresponds to the average value of the von Mises
strain in the sample and $N$ is the total number of atoms. 
We have also computed the volumetric strain. Its magnitude, however, is two orders of magnitud smaller than the von Mises strain. Also, its behavior does not appear to shed additional light on the sample behavior so we will not discuss it further. The von Mises strain is the magnitude of the strain that is independent of volume change. As noted, it directly measures relative translations. We will consider a second descriptor that quantifies the rotation of an atom's neighborhood, for each atom. This is obtained from the deformation gradient tensor $F$ as constructed by Zimmerman \emph{et al}~\cite{Zimmerman2009}, for atomic systems.  $F$ can be descomposed into the matrix product $F = RU$ of a rotation $R$ and a stretch $U$~\cite{Tucker2011}. The per--atom angle of rotation is calculated using the OVITO software~\cite{Stukowski2010}. Here we adopt a similar formula for the rotation angle localization, $\phi$, which we define as 
\begin{equation}
\phi = \sqrt{\frac1N \sum_{i=1}^N\left(\theta_i - \theta_{\text{ave}}\right)^2}\ ,
\label{eq:angle}
\end{equation}
where $N$ is the number of atoms, $\theta_i$ corresponds to the rotation angle for atom $i$ and $\theta_{\text{ave}}$ to the average of the rotation angle for all the atoms in the sample.

An important descriptor of amorphous material behavior at the  mesoscopic scale is the non--affine displacement. That is, displacements  that deviate from displacements that can be described by a linear strain field (see for example~\cite{Langer1998} and~\cite{Tanguy2002,Tanguy2006}). We shall focus on non--affine displacements, and will introduce a global descriptor thereof, the mean non--affine square displacement of the sample,
\begin{equation}\label{r2na}
\mathcal{R}^2_{\text{na}} \equiv \frac1N\sum^N_{i=1}\left|\vec{r}_{\text{na}}^{\;i}\right|^2\; ,
\end{equation}
where $\vec{r}_{\text{na}}^{\;i}$ corresponds to the non--affine displacement vector of the
$i-th$ particle. This descriptor will be useful to characterize the strain at which shear bands are formed.

From an atomic structure point of view, it will be of interest  to correlate changes in the non--affine displacement with the packing of the atomic structure. To do this, we shall use a Voronoi analysis~\cite{Jonsson1988}, where we characterize each atom with its Voronoi 
indexes $\left<n_3, n_4, n_5, n_6\right>$. With these Voronoi indices it is possible to 
define the local fivefold symmetry (L5FS)  (Hu~\emph{et al.}~\cite{Hu2015})
as the number of pentagons in each polyhedron over the total number of faces that 
constitute it, that is
\begin{equation}
 \text{L5FS} = \frac{n_5}{\sum_i n_i} \ .
\end{equation}
This descriptor allows us to evaluate the stability of the atomic structure in MGs, as 
was previously described in~\cite{Wang2014,Zhi2017}. 

\section{Results}
\begin{figure}[h]\centering
\includegraphics[height=6cm]{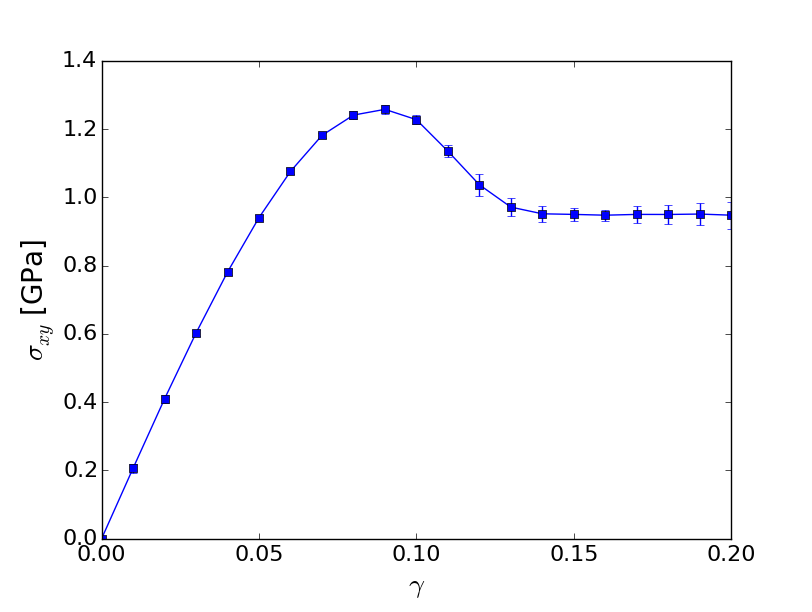}
\caption{Averaged stress--strain curve of ten samples under simple shear deformation test. Error bars indicate the standard deviation. Non-affine displacement (NAD) begins to localize in homogeneous bands at about $\gamma \sim 0.7$. Shear bands (SBs) begin nucleating at about $\gamma \sim 0.10$. Failure occurs at $\gamma = 0.14$. See text for details.}
\label{ss-10seed}
\end{figure}
The ten samples were loaded in simple shear up to $\gamma = 0.20$ as described in 
the previous section. Figure \ref{ss-10seed} shows the average stress--strain curve. 
There is linear elastic behavior only for small strains, up to $\gamma \sim 0.04$, 
with a shear modulus of $G=21.43$~GPa. Thereafter the curve bends downwards until 
reaching a maximum at $\gamma \sim 0.08$ with an ultimate tensile strength of 
$\sim 1.25$~GPa. There is little sample-to-sample variation as indicated by the small standard deviations. This behavior is in agreement with current understanding of amorphous materials: isolated STZs
are generated at small strains~\cite{Spaepen1977,Argon1979}, and they provide an accumulating  deviation from linear elasticity. At about  $\gamma \sim 0.07-0.08$ serious plastic deformation sets in, and we shall focus on what happens within this region.

\begin{figure}[]\centering
\includegraphics[height=6cm]{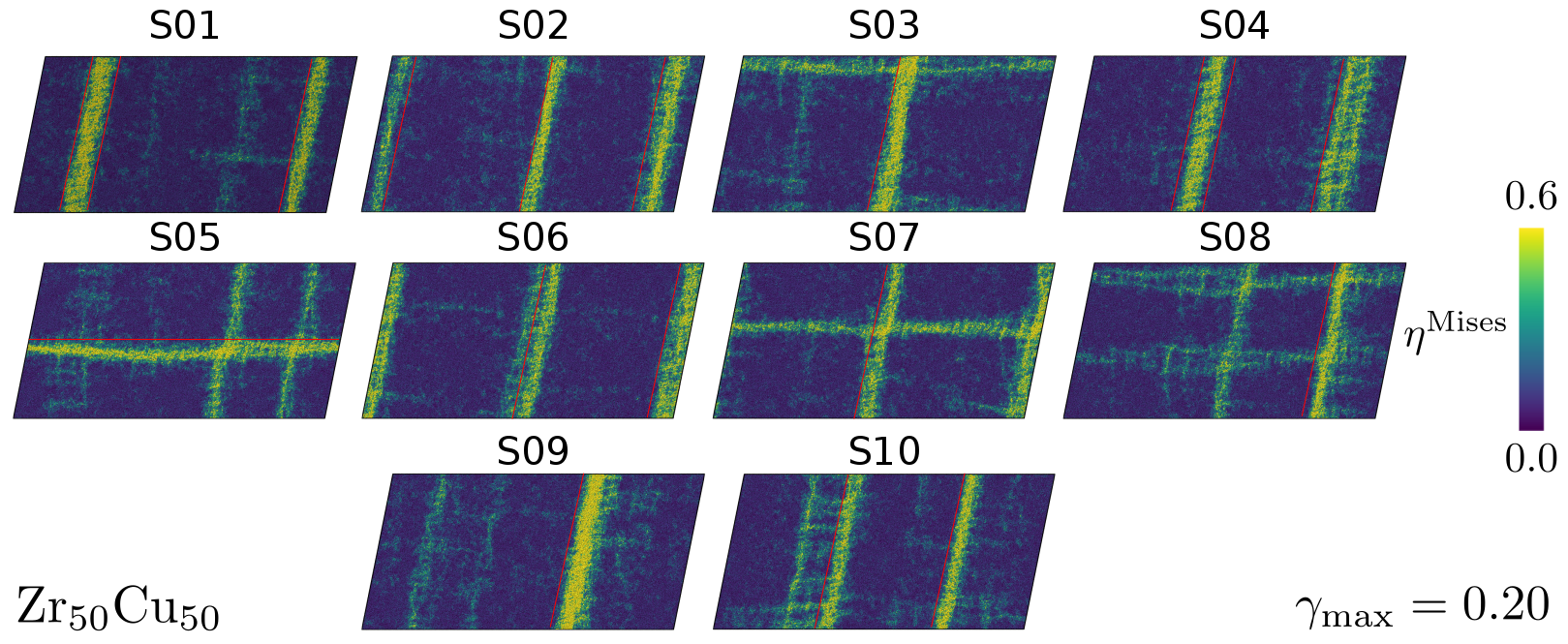}
\caption{Von Mises strain $\eta^{\text{Mises}}$ at a strain of $\gamma=0.20$ for the ten Cu$_{50}$Zr$_{50}$ samples. Color indicates von Mises value. At $\gamma \approx 0.14$ all ten samples fail along one, two, or three surfaces (``fault planes''---FPs) located at the boundaries of shear bands, indicated  by red lines. Samples S03, S07, S08 and S09 have one FP, samples S01, S04, S06 and S10 have two FPs and S02 and S05 have three FPs.}
\label{vonmisesXL}
\end{figure}

At about $\gamma \sim 0.12$ 
shear bands are fully identifiable, and the material fails at $\gamma =0.138\pm0.006$. This is noticed as a 
sliding of the blocks adjacent to the SB along a ``fault plane''. Although failure occurs at the same applied 
strain in all ten samples, the actual mechanism is different in each sample: It can occur along one fault plane 
only, or simultaneously along two or three fault planes. These fault planes coincide with boundaries of shear 
bands. After failure the numerical simulation can be continued due to the periodic boundary conditions.   
Here we defined a SB as regions with values of the von Mises strain above 0.3 and that completely cross the samples.
Figure~\ref{vonmisesXL} shows the ten samples at $\gamma_{\text{max}} = 0.20$ where the 
atoms are coloured according to their local von Mises strain, where the minimum is for 
purple at $\gamma =0.0$ and the maximum for yellow at $\gamma = 0.6$. It is apparent that, although the global 
stress--strain behavior of the different samples is the same, their detailed evolution 
at the mesoscopic level is quite different: Strain is, as expected, localized in 
shear bands, as measured by the von Mises strain. There can be one, two or three bands, 
and they can point in any of the two directions of maximum shear stress.
n
\subsection{Local analysis}
\begin{figure}[h!]\centering
\includegraphics[height=7.2cm]{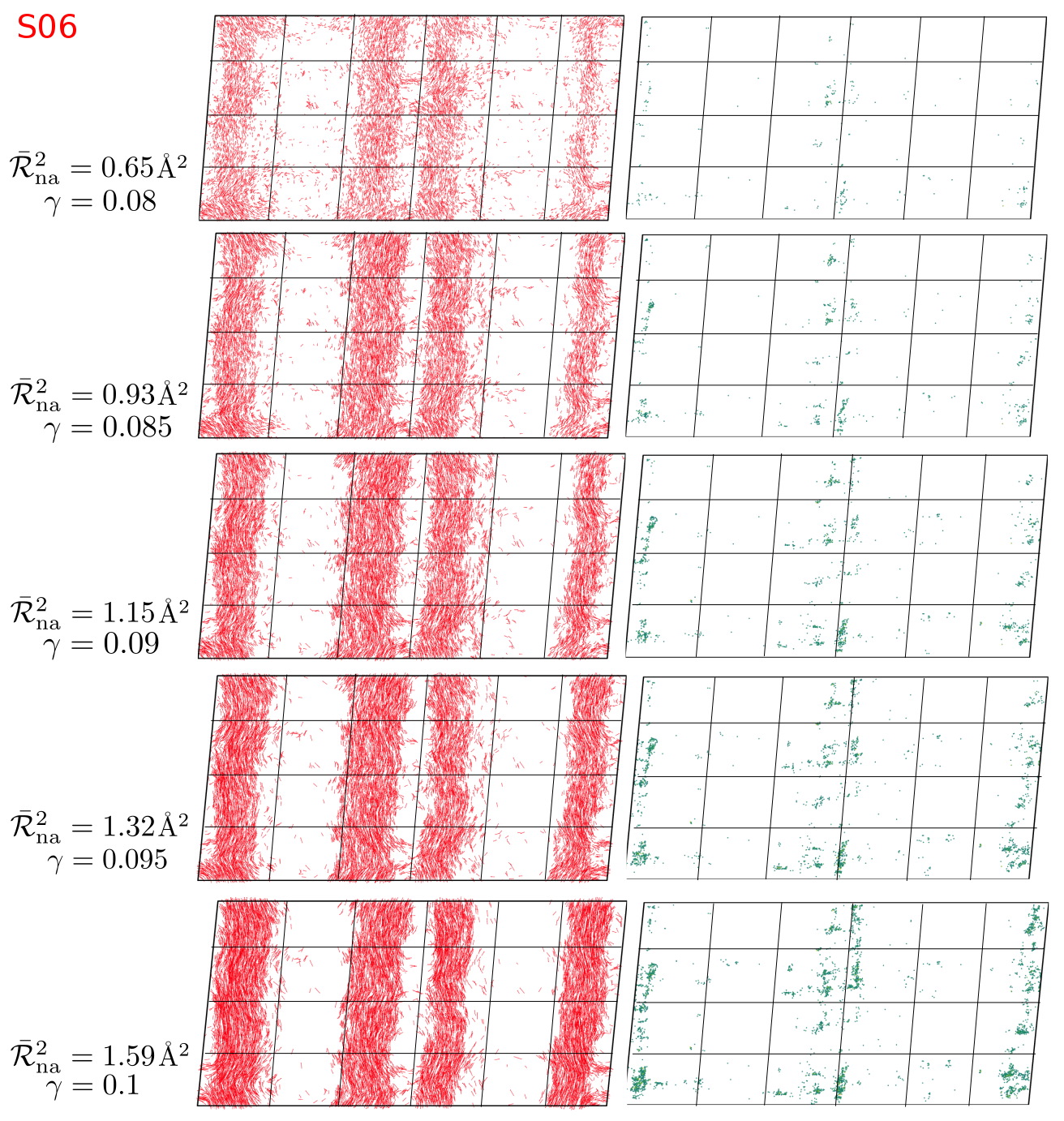}
\includegraphics[height=7.2cm]{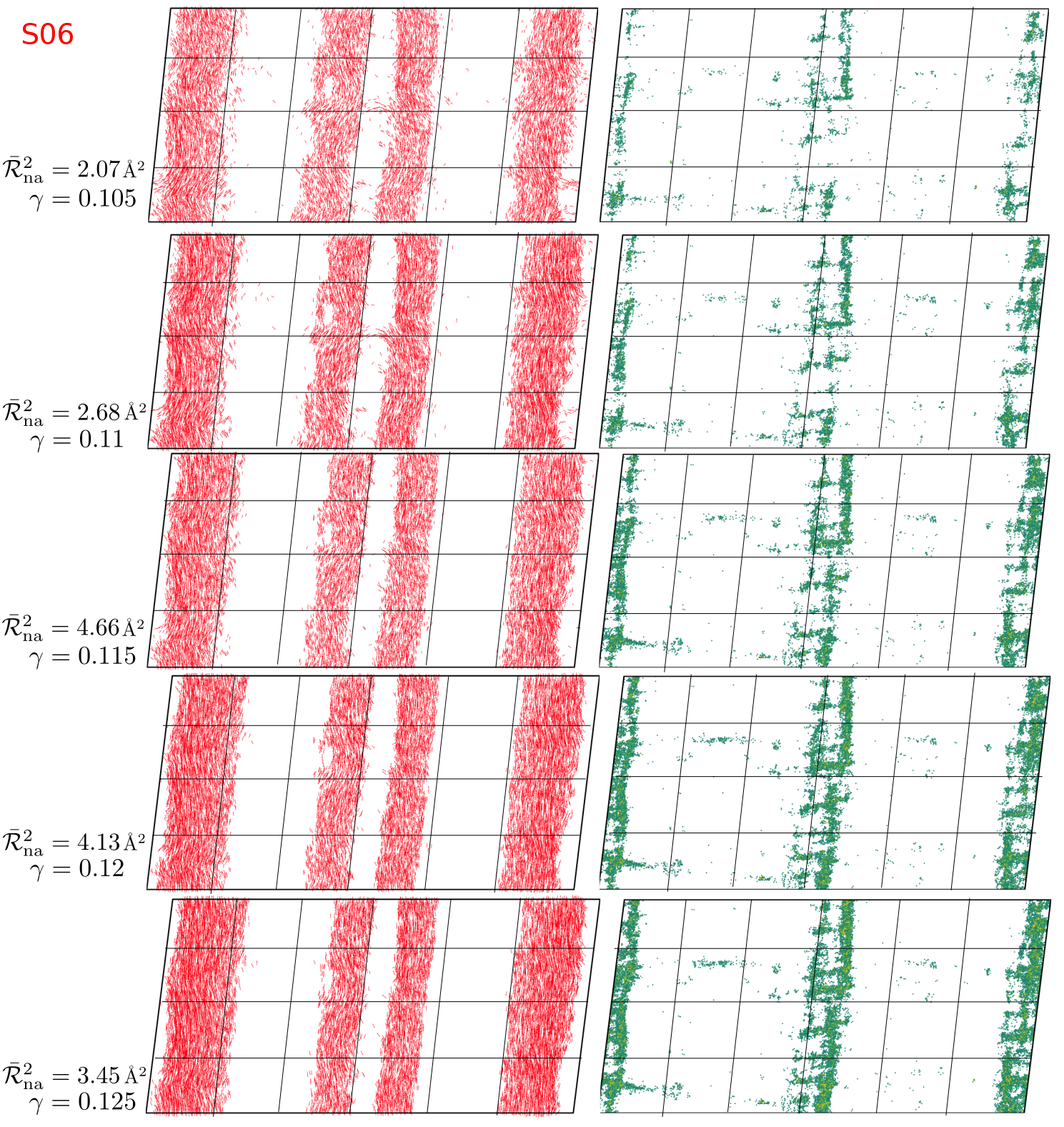}
\caption{Behavior of non--affine displacement (NAD) and local von Mises strain in the region of plastic deformation leading up to failure. The left-hand panels show atoms with NAD bigger than average, whose value is indicated. The right-hand panels show atoms with von Mises strain greater than 0.3. NAD is localized along bounded regions well before shear banding starts. The latter activity does not overlap with the regions of high NAD. A grid is drawn to help visualized this NAD-SB avoidance. Shown here are the results for sample S06. All other samples behave in a similar fashion (not shown).}
\label{evolve}
\end{figure}

Figure~\ref{evolve} shows the evolution of both the non--affine displacement field and the 
von Mises strain of one sample in the whole material when the external shear $\gamma$ changes 
from 0.8 to 0.125. This is the strain interval where shear bands form and leads up to failure. 
In accordance with \cite{Langer1998} we define the non--affine part of the displacement due to the external shear 
deformation as $\vec{r}_{\text{na}} = x_{\text{na}}\hat{x} + y_{\text{na}}\hat{y}$, where $x_{\text{na}} = x_i - \left(y_i/\dot\gamma\right)t - x_0$,
$y_{\text{na}} = y_i - y_0$ and where $\dot\gamma$ is the shear rate, \{$x_i$, $y_i$\} the current
coordinates of the particle $i$ and the reference coordinates \{$x_0,y_0$\}. 
For the non--affine displacement field, shown as red arrows, we only display atoms whose NAD is above the mean at each 
stage of deformation. On the right panel we show those atoms that exhibit a von Mises strain $\eta^{\text{Mises}} \geq 0.3$. 

From the behavior of NAD and SB displayed on Figure~\ref{evolve} the following 
scenario emerges: First of all, the NAD organizes itself along narrow stripes that 
follow the directions of maximum resolved shear stress, and pointing along said 
directions, well before there is even a hint of shear banding. It is reasonable that 
NAD organizes in regions where it points roughly along the same direction, in order 
to minimize strain and hence to minimize energy. As external strain is increased, 
NAD is increased as well, and the displacement gradients, i.e., the internal stresses 
increase {\it at the edges} of the regions of high NAD. It is these extended regions 
that play the role of the notch in the numerical experiment of~\cite{Sopu2017}.  
This generates a front, and all along this front regions of high von Mises strain, 
i.e., SBs, are formed. This scenario holds in the remaining nine samples as well (not shown).

\begin{figure}[h!]\centering
\includegraphics[height=6cm]{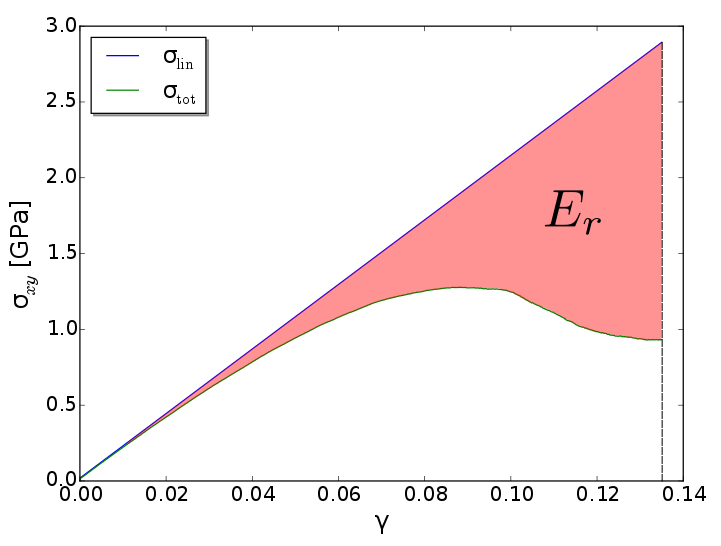}
\caption{Difference between the area under the linear regime of stres--strain curve ($\sigma_{\text{lin}}$)
and the total stress--strain curve ($\sigma_{\text{tot}}$) marked as $E_r$.This energy is absorbed by the shear bands and roughly determines their thickness.}
\label{SS_Er}
\end{figure}

The SB growth is limited because they consume a large amount 
of inelastic energy, so their thickness is limited. To verify this conclusion we 
perform an energy analysis of all the samples. First, as we perform the deformation of the 
system using the NVE ensemble, there exists an increase, on average, of the temperature up 
to $T\approx57$~K. This increase only uses up about 1\% of the available energy due to the externally 
applied stress. Second, to quantify the energy used to form the shear bands, we take the 
difference between the area under the stress--strain curve 
of the projected elastic behavior and the total stress--strain curve up to $\gamma=0.135$, as can be seen in 
figure~\ref{SS_Er}. This will give us an energy per unit volume 
of each system under study. We compare this quantity with the energy per unit volume of the atoms 
that belong to the shear bands, defined as:
\begin{equation}
E_r = \frac12\sum_{i\in \text{SBs}}^N \sigma_{xy}^i\cdot\eta^{xy}_{i}\; ,
\end{equation}
where $\sigma_{xy}^i$ are the components of the per--particle stress tensor of the $i-th$ atom, $\eta^{xy}_i$ the 
components of the per--particle strain tensor of the $i-th$ atom in the SBs. Here the Einstein sum rule is applied over the $x,y$ indices.
Thus, $E_r$ will correspond to the energy used by the system to form the shear bands. As a result, taking the difference between 
the potential energy of the system and $E_r$ we obtain in average $0.063\pm0.007$~eV~\AA$^{-3}$. Comparing with the result from the 
area under the stress--strain curve $E_r$, of $0.065$~eV~\AA$^{-3}$, we have a difference at the 3\% level. This result supports our previous argument: the SBs consume a large amount of inelastic energy which limits their thickness. The average width of the shear bands of our simulations is $\sim 40$ \AA.

\begin{figure}[h!]\centering
\includegraphics[height=7.2cm]{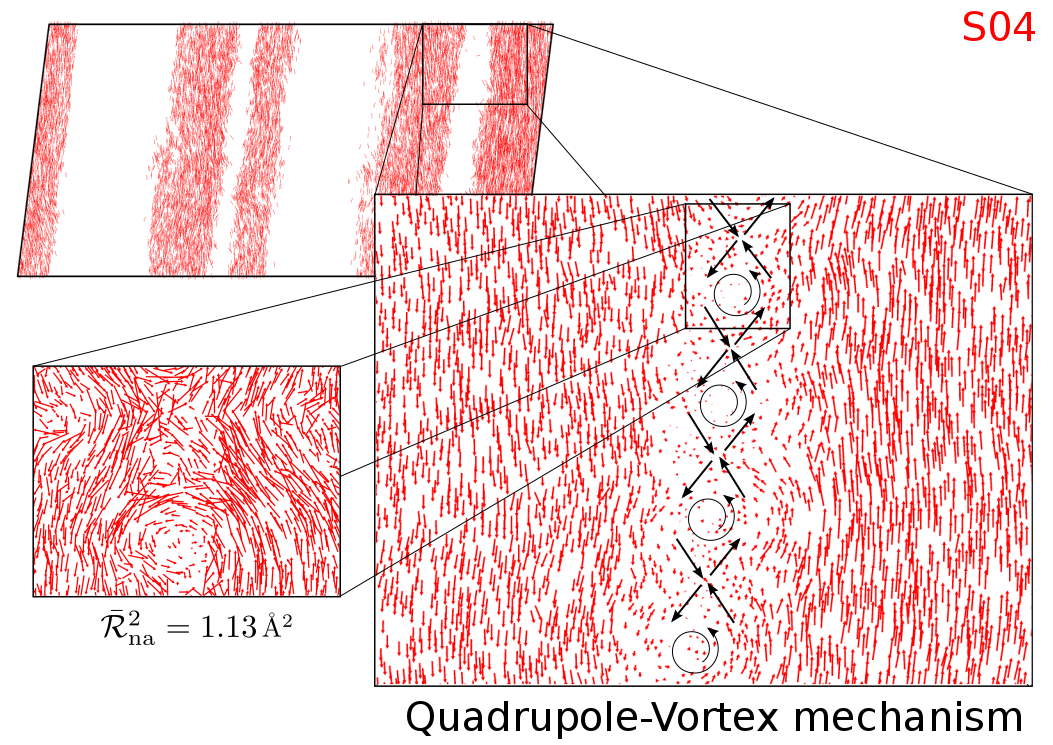}
\caption{NAD for sample S04, at $\gamma=0.10$, where only atoms above the average are shown. The 
zoom corresponds to a complete NAD field, with alternating quadrupole--vortex regions. The same behavior obtains for all SBs in all samples (not shown).}
\label{SV}
\end{figure}

\begin{figure}[h!]\centering
\includegraphics[height=5cm]{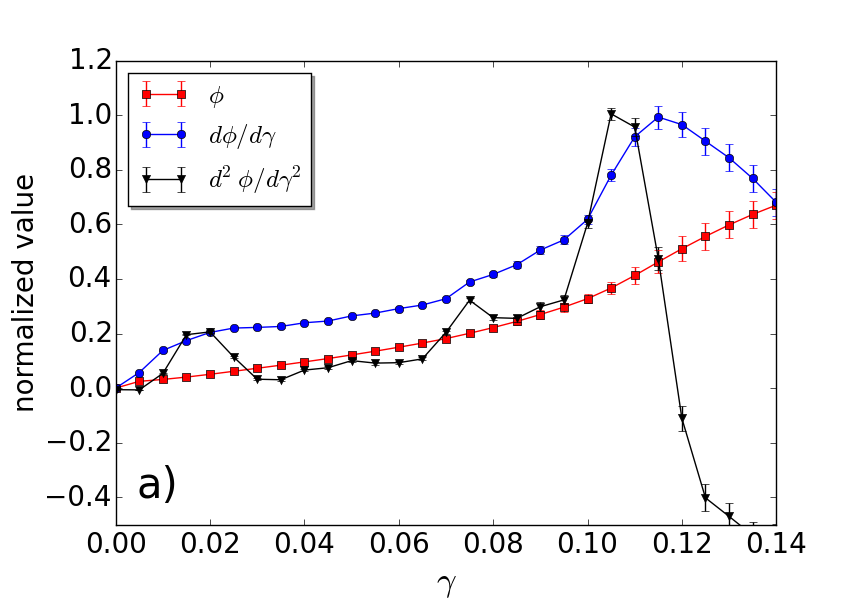}
\includegraphics[height=5cm]{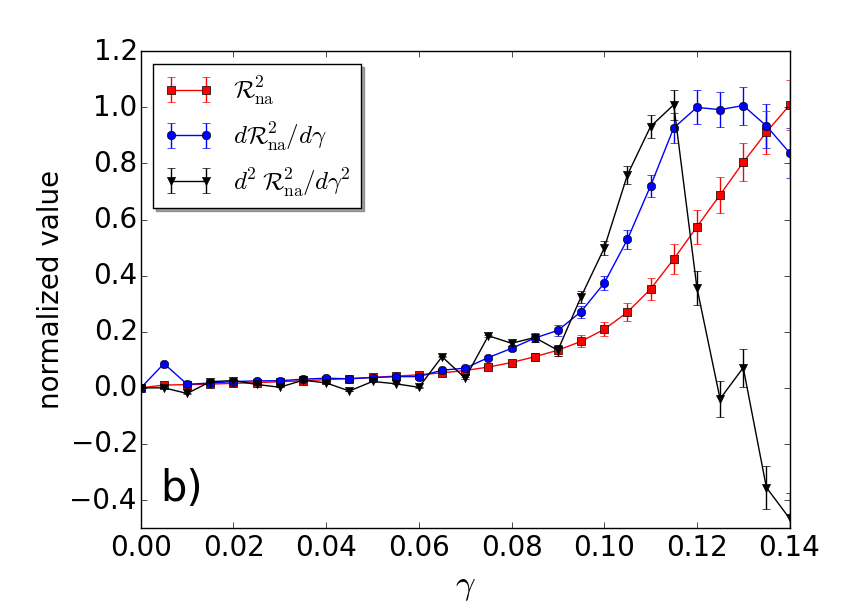}
\includegraphics[height=5cm]{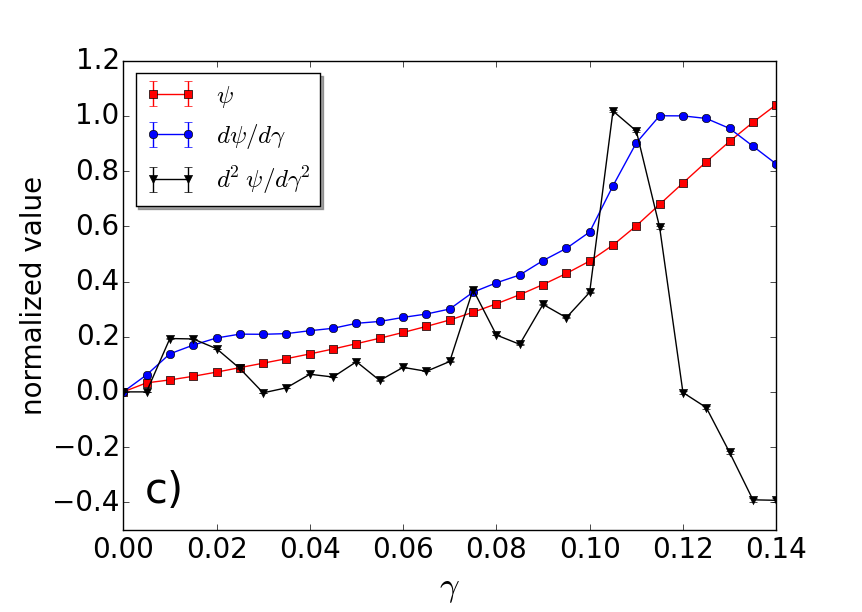}
\caption{Normalized value of (a) averaged rotation angle localization $\phi$, (b) NAD $\mathcal{R}^2_{\text{na}}$ and (c) degree of strain localization $\psi$ with thier first and second derivatives as a function of macroscopic deformation $\gamma$.In all three cases, there is a sharp acceleration that coincides with the onset of shear banding at $\gamma \sim 0.10$. Each quantity is first computed as an average over a single sample. The result is then averaged over the ten samples. Error bars are the standard deviations in the latter averaging.}
\label{fig: DDS}
\end{figure}

As a result, SBs and regions of high NAD ``avoid each other'' and give rise to rotational motion 
in regions that comprise the SBs, as presented in figure~\ref{SV}. Here it is possible to 
identify a mechanism of deformation similar to the mechanism  presented by \c{S}opu \emph{et al}.~\cite{Sopu2017}:
Within a SB, NAD is low, and there are regions with high rotational motion that alternate 
with regions of quadrupole--like displacement. In our case there is no externally generated 
flaw, so this alternation occurs, rather than sequentially along a propagating SB, homogeneously 
along the region occupied by the SB. To quantify this rotational motion we have calculated the rotation angle localization $\phi$
averaged over the 10 samples as well as its first and second derivative with respect to macroscopic strain $\gamma$. 
This is presented in figure~\ref{fig: DDS}(a). At $\gamma=0.10$, the onset of shear banding, there is 
a sudden, and significant (by a factor of four) acceleration in the rotation of the particles, 
as measured by the second derivative  $d^2\phi/d\gamma^2$. This is followed by a deceleration of the 
rotation when the SBs are already consolidated at failure $\gamma=0.14$. 

\subsection{Global Analysis}

Figure~\ref{fig: DDS} shows, in addition to the behavior of the rotation angle localization $\phi$, two global parameters, the mean non--affine square displacement $\mathcal{R}^2_{\text{na}}$, Eq. (\ref{r2na}), together with its first and second derivative, and the degree of strain localization $\psi$, Eq. (\ref{eq:psi}), with its first and second derivative as well, as a function of external strain $\gamma$. All three quantities show an unmistakable acceleration, at $\gamma \sim 0.10$, when the shear banding sets in.

\begin{figure}[h!]\centering
\includegraphics[height=6cm]{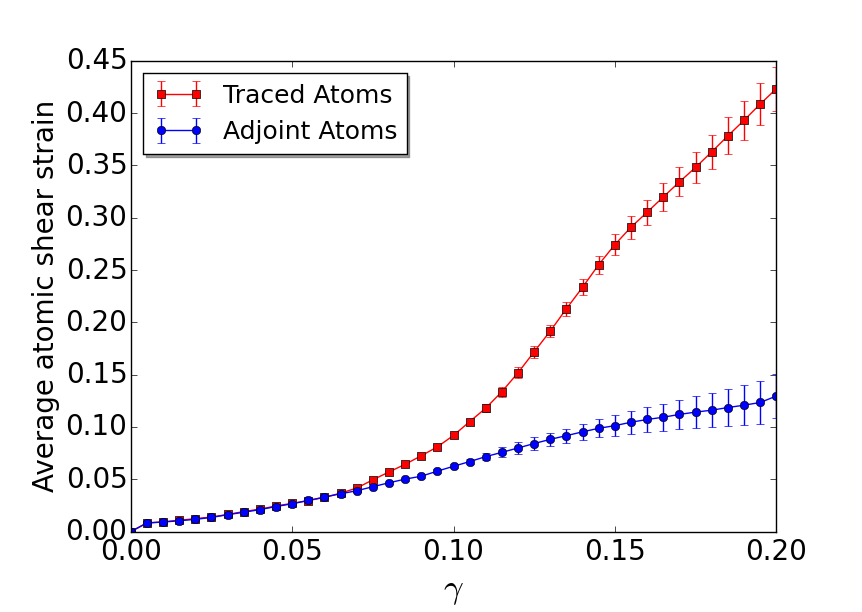}
\caption{Average atomic shear strain for traced atoms, defined as atoms that experience a local von Mises strain grater than 0.3 after the shear band formation. Adjoint atoms represent the neighbors of traced atoms after the shear band is already formed. There is a sharp discontinuity in shear strain between the highly strained atoms (mostly belonging to shear bands) and their neighbors. Notice the accelerated rate of increase in shear strain from the beginning of plastic behavior ($\gamma \sim 0.7$) up to failure at $\gamma = 0.14$.}
\label{tracedvsadjoint}
\end{figure}

\begin{figure}[h!]\centering
\includegraphics[height=6cm]{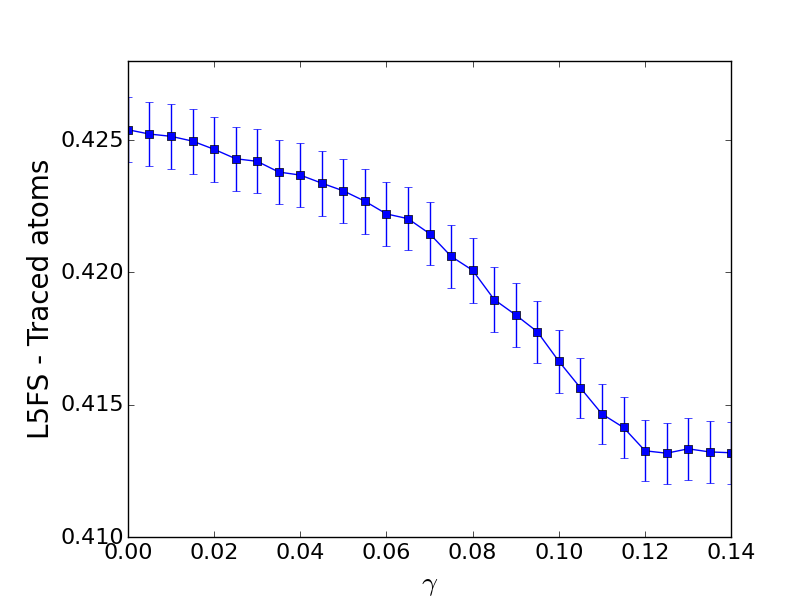}
\caption{Averaged local five--fold symmetry (L5FS) for traced atoms as a function of macroscopic deformation. There is a steady decrease in the L5FS, with a slight increase in slope at the onset of plastic behavior at $\gamma \sim 0.7$, until shear bands are, roughly, fully formed at $\gamma \sim 0.12$. L5FS is then stabilized until failure at $\gamma = 0.14$.}
\label{l5fs-traced}
\end{figure}

Further evidence that high stress is concentrated in localized patches is obtained focussing on atoms that, at the end of the simulation run, have $\eta^{\text{Mises}} \geq 0.3$. Following~\cite{Zhi2017} we call them ``traced atoms'' and we compare the behavior of their number as a function of strain $\gamma$ with the behavior of their neighbors, called ``adjoint atoms''. The result of this exercise is presented in figure~\ref{tracedvsadjoint}, where the traced atoms show a change in the rate of increase of the atomic shear strain around 
$\gamma \sim 0.10$, similarly as the non--affine displacement field, $\mathcal{R}_{\text{na}}^2$, the rotation angle and the degree of strain localization $\psi$. As shown in the figure~\ref{fig: DDS}. The adjoint atoms, the neighbors to the highly strained ones, do not show this sharp increase in the second derivative. 

Finally, the local five--fold symmetry has been linked in previous works, to the microscopic beahavior at the glass transition~\cite{Jonsson1988,Hu2015} and within SBs~\cite{Tercini2018}.
In order to clarify the local structure of the shear band, we will focus our attention on the traced atoms and, as an 
indication of the local structure, we will calculate the local five--fold symmetry, L5FS, for those atoms. The results 
of L5FS for the traced atoms during the shear deformation tests are presented in the figure~\ref{l5fs-traced}. The results 
of figure~\ref{l5fs-traced} indicate that the pentagonal structure of the traced atoms decreases during the formation of 
the shear band, and then stabilizes. The decrease, however, appears to be monotonous,  although there appears to be some acceleration in the decrease of L5FS at $\gamma \sim 0.7$. 

\section{Discussion and Conclusions}

Applying a uniform shear to ten samples of Cu$_{50}$Zr$_{50}$ MG we have found some interesting common features: First, their global behavior is very similar but their behavior at the atomic scale is very different, as they all fail at the same external strain, but their specific mechanism of failure differs from sample to sample. It can be along one, two, or three fault planes, that are at the boundary of shear bands. Second, there is a self-organization of non--affine displacement along localized bands, and this self-organization precedes and apparently triggers, the birth of shear bands. This fact, if it persists under conditions that differ from the conditions we have considered in the present numerical experiment, and in the absence of external flaws, could be an interesting candidate to play the role of precursor, to be closely monitored, to the nucleation of shear bands and subsequent failure. Third, the second derivative of several quantities (rotation angle localization, strain localization and non--affine square displacement) shows a sharp increase at onset of shear banding, preceding failure.

Our simulation has been carried out in shear, on a thin slab of Cu$_{50}$Zr$_{50}$ metallic  glass at a temperature of 10 K, and at externally imposed strain rate of $\dot\gamma=5\times10^8$~s$^{-1}$.  It will be of interest to see if our conclusions hold for other values of the driving parameters, particularly temperature and strain rates. For example, Albe et {\it al.}~\cite{Albe2013} have used higher and lower strain rates for a tension test of Cu$_{64}$Zr$_{36}$ at 50 K and 300 K, and have uncovered a rich overall stress--strain behavior which it would be interesting to correlate with atomic--scale behavior.

\section*{Acknowledgements}

We gratefully acknowledge the support of the ECOS-Conicyt program. The work of FL has been supported by Fondecyt Grant 1191179.
GG thanks Fondecyt Grant 1171127.





\end{document}